\documentclass[twocolumn,english,aps,prb,twocolum,superscriptaddress,nobibnotes,amsmath,amssymb,floatfix]{revtex4-1}
\usepackage[utf8]{inputenc}
\usepackage[english]{babel}
\usepackage{amsmath,amsfonts,amssymb}
\usepackage{upgreek}
\usepackage[T1]{fontenc}
\usepackage{url}
\usepackage{placeins}

\usepackage[version=3]{mhchem}

\usepackage{bibunits}
\defaultbibliography{Biblio}
\defaultbibliographystyle{naturemag}
\usepackage{epstopdf}
\usepackage{graphicx}
\graphicspath{{./Figures/}}

\usepackage{titlesec}
\titleformat{\section}{\normalfont\large\bfseries\filcenter}{\thesection}{1em}{}
\titlespacing{\section}{0pt}{10pt}{5pt}

\newcommand{\Fref}[1]{Figure~\ref{#1}}
\newcommand{\fref}[1]{Fig.~\ref{#1}}

\def \DWRep {\Delta \omega_{\rm rep}}
\def \DFRep {\Delta f_{\rm rep}}

\begin{document}

\title{Spatial multiplexing of soliton microcombs}

\author{E.~Lucas}
\affiliation{IPHYS, École Polytechnique Fédérale de Lausanne (EPFL), CH-1015 Lausanne, Switzerland}

\author{G.~Lihachev}
\affiliation{Russian Quantum Centre, 143025, Skolkovo, Russia}
\affiliation{Faculty of Physics, M.V. Lomonosov Moscow State University, 119991
Moscow, Russia}

\author{R.~Bouchand}
\affiliation{IPHYS, École Polytechnique Fédérale de Lausanne (EPFL), CH-1015 Lausanne, Switzerland}

\author{N.~G.~Pavlov}
\affiliation{Russian Quantum Centre, 143025, Skolkovo, Russia}
\affiliation{Moscow Institute of Physics and Technology, 141700, Dolgoprudny, Russia}

\author{A.~S.~Raja}
\affiliation{IPHYS, École Polytechnique Fédérale de Lausanne (EPFL), CH-1015 Lausanne, Switzerland}

\author{M.~Karpov}
\affiliation{IPHYS, École Polytechnique Fédérale de Lausanne (EPFL), CH-1015 Lausanne, Switzerland}

\author{M.~L.~Gorodetsky}
\affiliation{Russian Quantum Centre, 143025, Skolkovo, Russia}

\affiliation{Faculty of Physics, M.V. Lomonosov Moscow State University, 119991
Moscow, Russia}

\author{T.~J.~Kippenberg}
\email[]{tobias.kippenberg@epfl.ch}
\affiliation{IPHYS, École Polytechnique Fédérale de Lausanne (EPFL), CH-1015 Lausanne, Switzerland}

%\begin{abstract}
%\end{abstract}
\maketitle

\begin{bibunit}

\noindent\textbf{\noindent
Dual-comb interferometry utilizes two optical frequency combs to map the optical field’s spectrum to a radio-frequency signal without using moving parts, allowing improved speed and accuracy.
However, the method is compounded by the complexity and demanding stability associated with operating multiple laser frequency combs.
To overcome these challenges, we demonstrate simultaneous generation of multiple frequency combs from a single optical microresonator and a single continuous-wave laser.
Similar to space-division multiplexing, we generate several dissipative Kerr soliton states -- circulating solitonic pulses driven by a continuous-wave laser -- in different spatial (or polarization) modes of a \ce{MgF2} microresonator.
Up to three distinct combs are produced simultaneously, featuring excellent mutual coherence and substantial repetition rate differences, useful for fast acquisition and efficient rejection of soliton intermodulation products.
Dual-comb spectroscopy with amplitude and phase retrieval, as well as optical sampling of a breathing soliton, is realised with the free-running system.
Compatibility with photonic-integrated resonators could enable the deployment of dual- and triple-comb-based methods to applications where they remained impractical with current technology.
}

%%%%%%%%%%%%%%%%%%%%%%%%%%%%%%
%%%%%%%%%%%%%%%%%%%%%%%%%%%%%%
%\section{Introduction}
%%%%%%%%%%%%%%%%%%%%%%%%%%%%%%
%%%%%%%%%%%%%%%%%%%%%%%%%%%%%%

% =============================================
\begin{figure*}[!t]
\centering
\includegraphics[width=.7\textwidth]{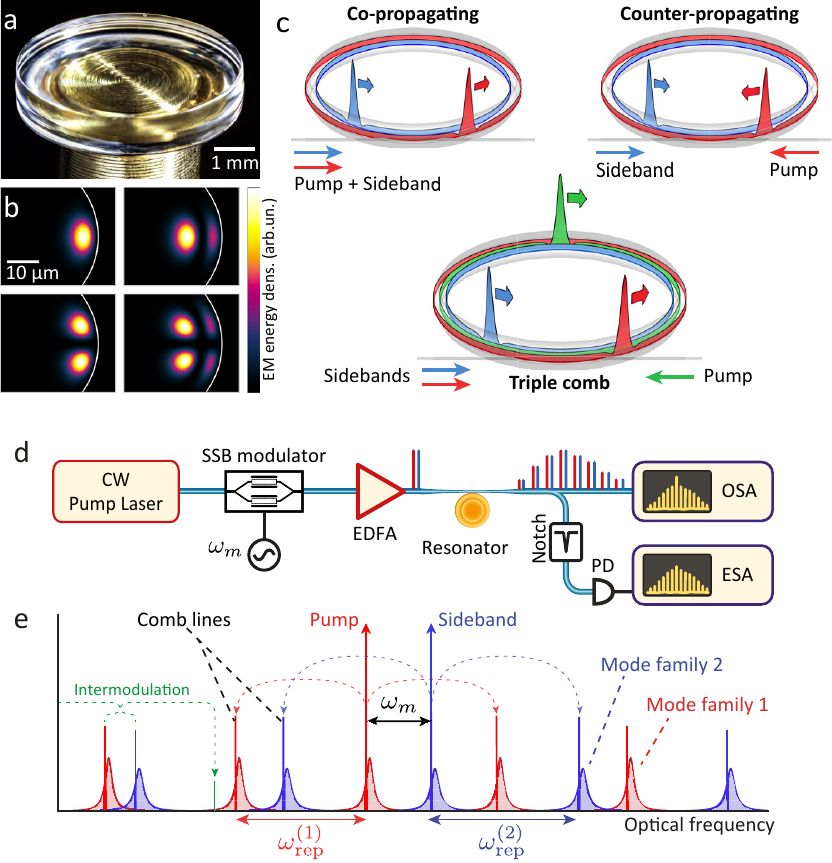}
\caption{\textbf{Principle of spatial multiplexing of solitons in a single microresonator}
\textbf{(a)} Crystalline \ce{MgF2} whispering gallery mode (WGM) resonator used in this work.
\textbf{(b)} Simulation of several optical mode profiles supported by the WGM protrusion.
\textbf{(c)} Schematic representation of the three schemes applied.
\textbf{(d)} Setup for dual DKS generation via spatial multiplexing in the co-propagating direction. The single-sideband (SSB) modulator creates an additional carrier to pump a second mode family. EDFA: erbium doped fibre amplifier. E/O-SA: electronic/optical spectrum analyser.
\textbf{(e)} Principle of multiplexed combs generation. The main pump laser (red arrow) is modulated to generate one sideband (blue arrow). The laser and sideband pump one resonance of two different mode families (1 in red and 2 in blue) and generate a soliton comb in each of them through the Kerr effect (red and blue lines). Co-propagating pulses may experience intermodulation effects (see SI for details).
}
\label{F1}
\end{figure*}
% =============================================

% =============================================
\begin{figure*}[!t]
\centering
\includegraphics[width=\textwidth]{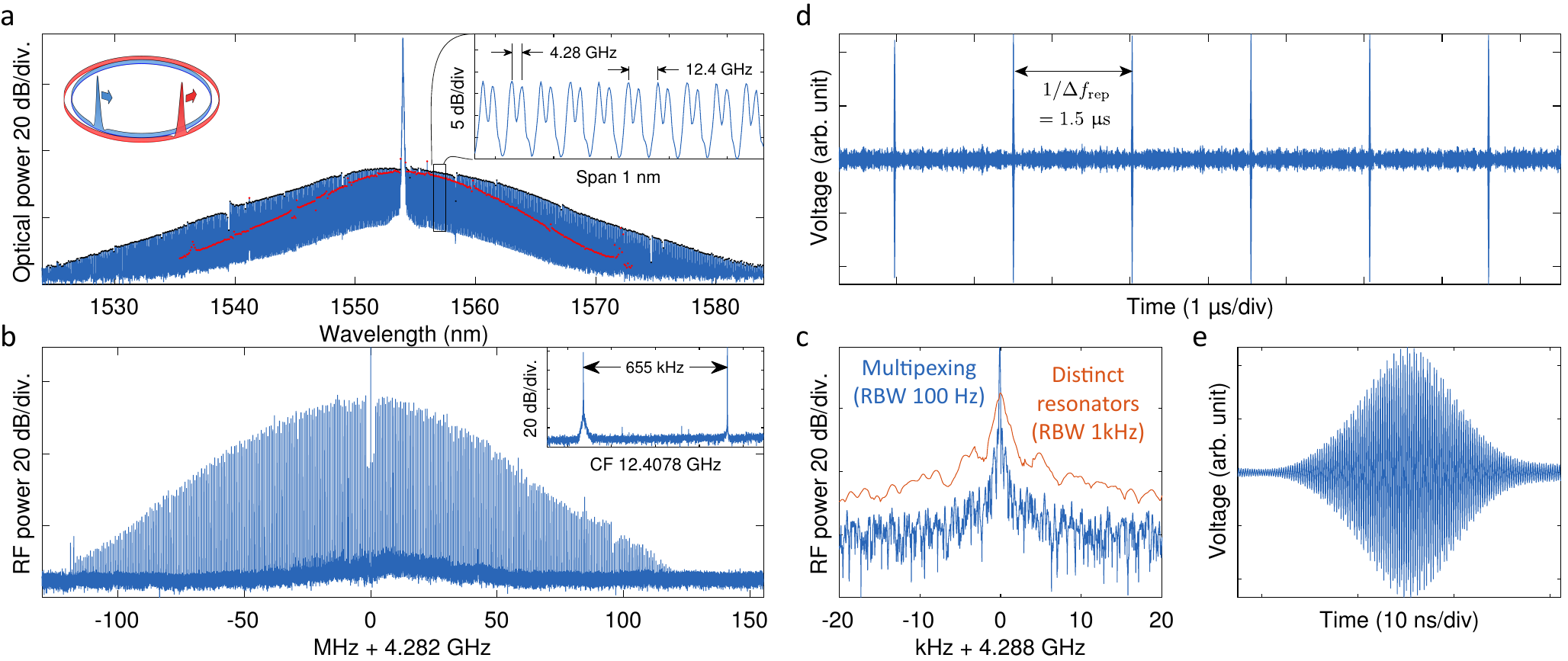}
\caption{\textbf{Dual-comb generation with spatially-multiplexed co-propagating solitons}
\textbf{(a)} Generated dual-comb optical spectrum. The DKS-based combs are interleaved and spaced by $\sim 4$~GHz (see inset). The red markers delineate one comb from the other.
\textbf{(b)} Resulting dual-comb RF heterodyne beatnotes. Resolution bandwidth (RBW): 3~kHz. The line spacing (repetition rate difference shown in inset) is 655~kHz. CF: centre frequency.
\textbf{(c)} Focus of one line of the RF comb. The blue trace denotes the multiplexed solitons in (b). The red represents the results from solitons generated in two distinct microresonators pumped with a single laser.
\textbf{(d)} Temporal interferogram of the dual-comb heterodyne shown in (b), recorded on a fast sampling oscilloscope, and after digital bandpass filtering to select the RF comb.
\textbf{(e)} Detail of the temporal trace (d) when the two pulses overlap ($\sim 200 \times$ magnification).
}
\label{F3}
\end{figure*}
% =============================================

% =============================================
\begin{figure*}[t]
\centering
\includegraphics[width=\textwidth]{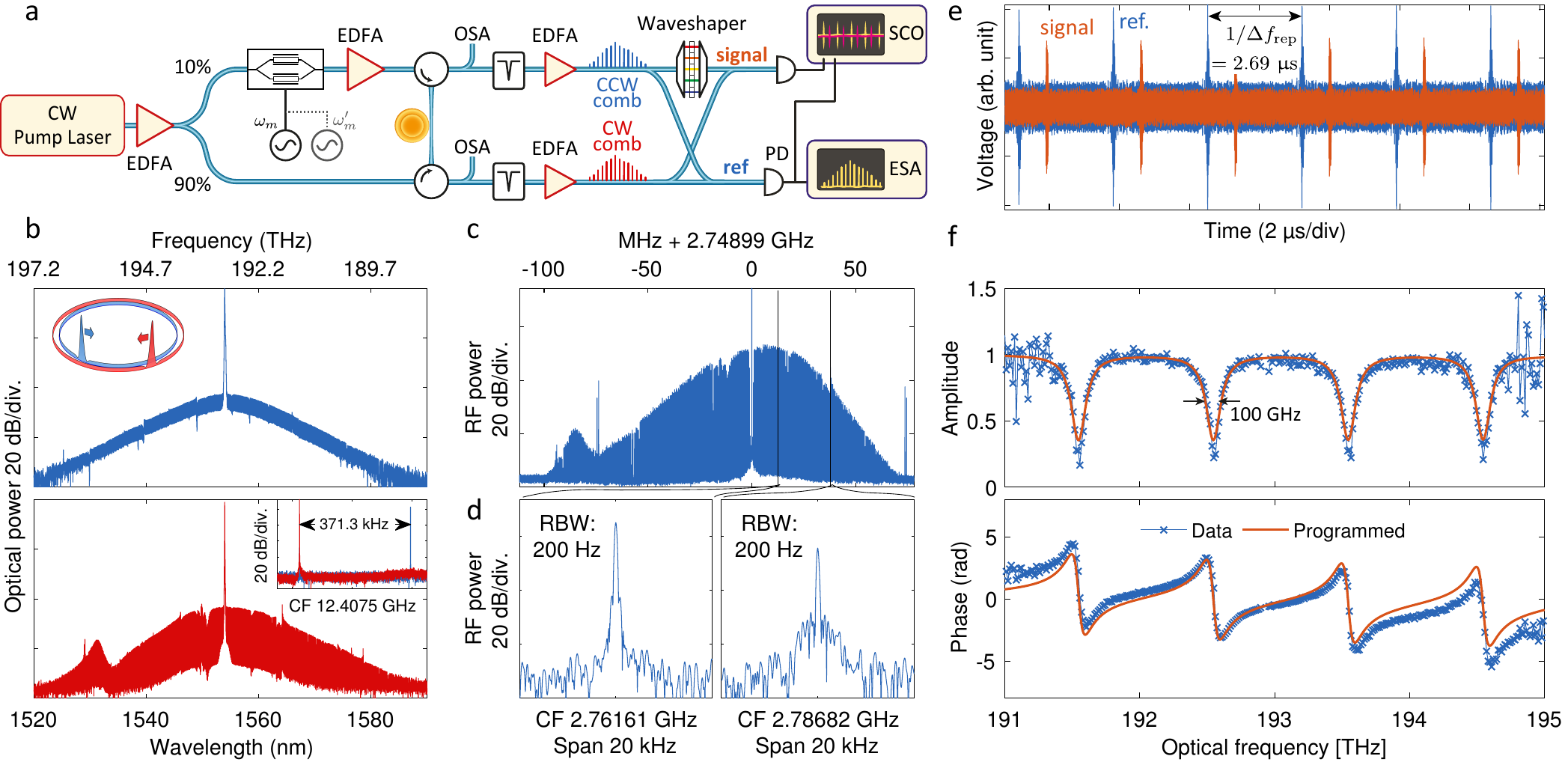}
\caption{\textbf{Dual-comb generation with spatially multiplexed counter-propagating solitons and proof-of-principle spectroscopy}
\textbf{(a)} Setup for counter-propagating dual and triple DKS-comb generation and spectroscopy.
\textbf{(b)} Optical spectra of the two counter-propagating combs. The inset shows the two repetition rate beats of the combs (CF: centre frequency).
\textbf{(c)} Resulting dual-comb beatnote (RBW 3~kHz).
\textbf{(d)} High resolution spectra of two lines of the RF comb in (c). 
\textbf{(e)} Temporal interferogram of the dual-comb heterodyne for the signal path (with waveshaper) and reference path.
\textbf{(f)} Retrieved amplitude and phase of the signal interferogram produced by coupling the dual-soliton pulse trains through a waveshaper programmed with synthetic absorption features (100~GHz FWHM). The orange lines display the programmed functions.
}
\label{F5}
\end{figure*}
% =============================================

% =============================================
\begin{figure*}[t]
\centering
\includegraphics[width=\textwidth]{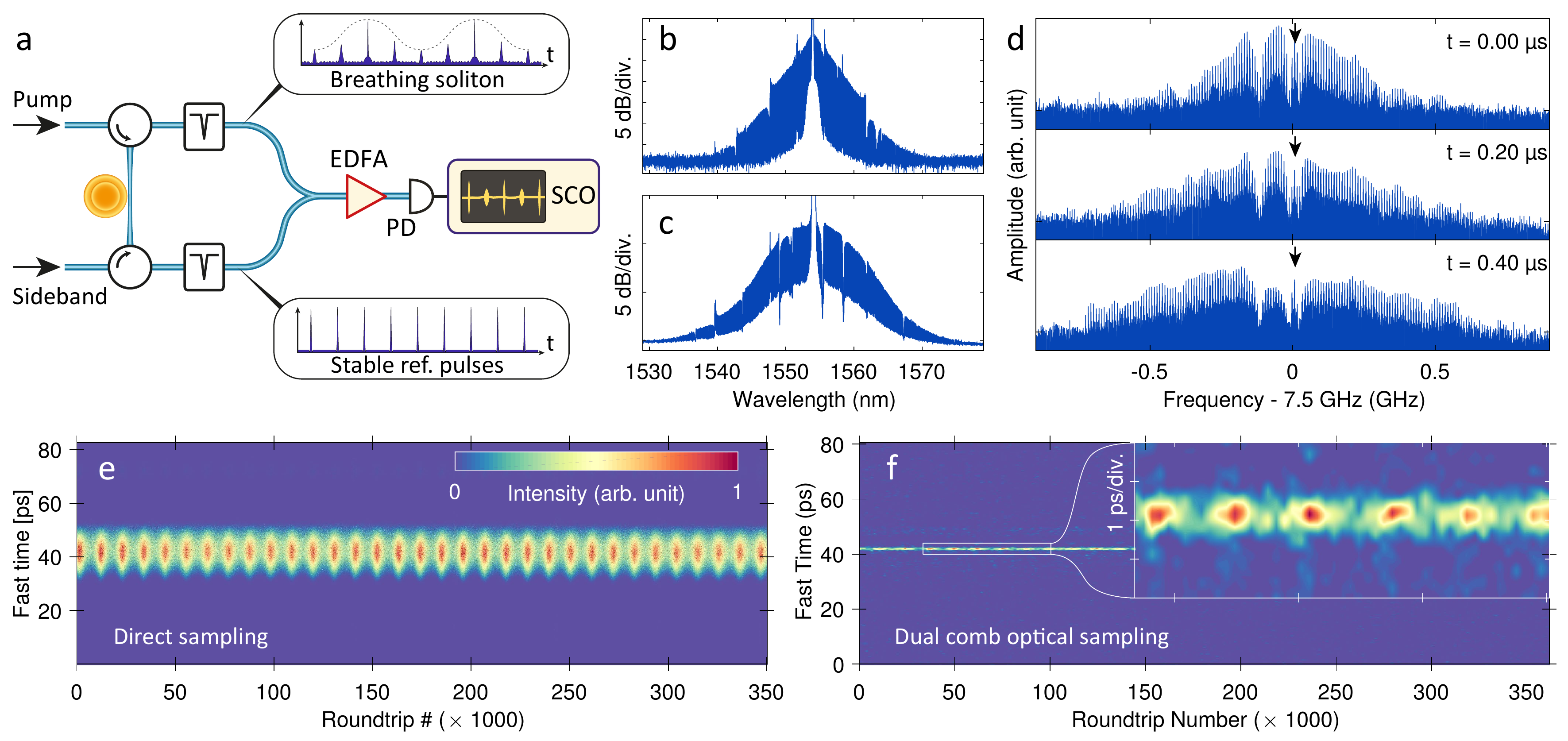}
\caption{
\textbf{Resolving the breathing dynamics of a soliton}
\textbf{(a)}~Experimental setup for breathing DKS dual comb imaging.
\textbf{(b)}~Optical spectrum of the breathing soliton pulse train and
\textbf{(c)}~of the stable reference pulse train.
\textbf{(d)}~RF spectrogram of the breathing soliton interferogram taken at maximum spectral contraction ($t=0~\upmu s$) and expansion ($t=0.4~\upmu s$). The arrow marks the pump position.
\textbf{(e)}~Spatiotemporal imaging of a breathing DKS via direct real-time sampling of the pulse train.
\textbf{(f)}~Same measurement realised with the multiplexed dual-comb. The fast time resolution is improved by an order of magnitude. The inset magnifies the white rectangular window.
}\label{F10}
\end{figure*}
% =============================================

% =============================================
\begin{figure*}[t]
\centering
\includegraphics[width=\textwidth]{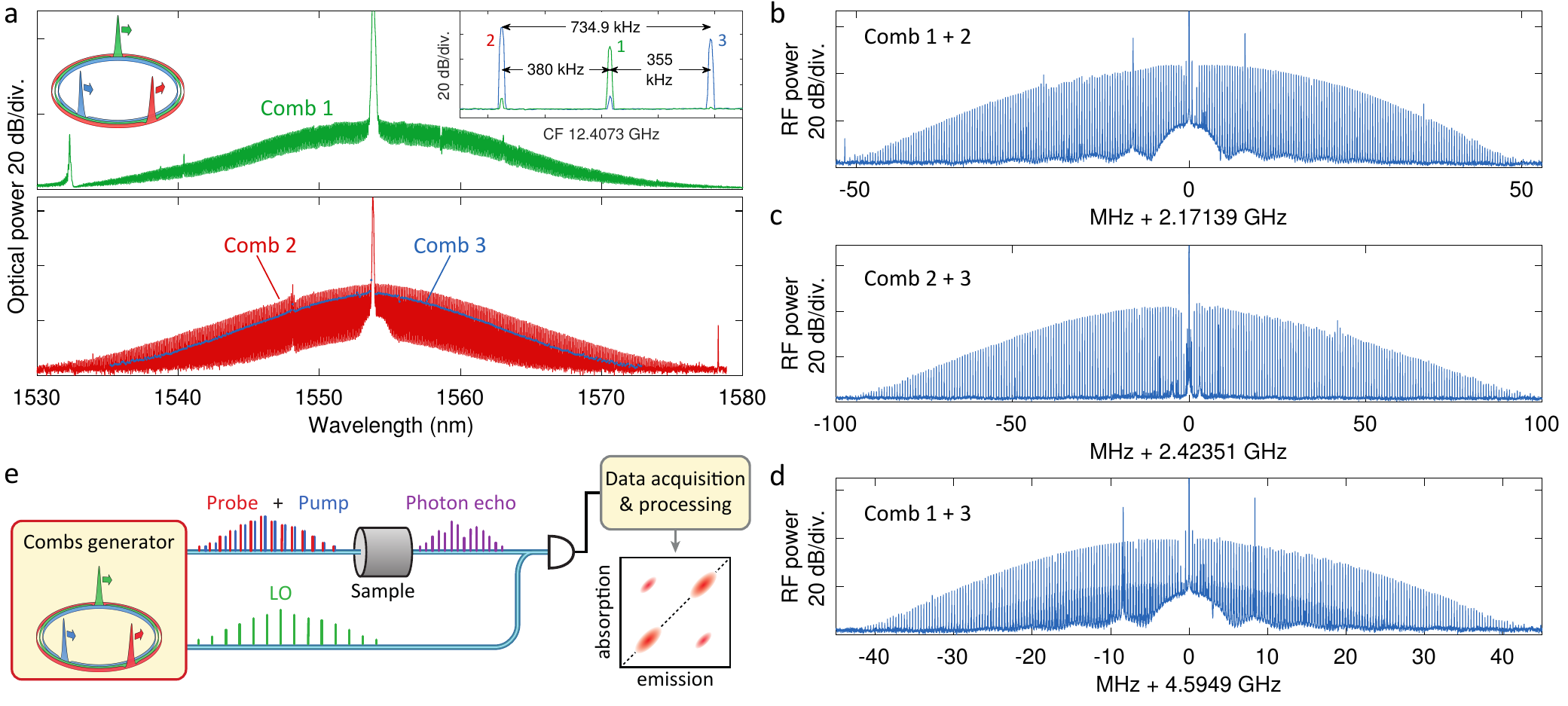}
\caption{\textbf{Triple comb generation in a single resonator by multiplexing in three mode families}
\textbf{(a)} Optical spectra. Comb~1 is generated in the CCW direction, while Comb~2 and 3 are generated in the CW direction. The inset shows the three distinct repetition rate beats.
\textbf{(b-d)} Heterodyning the three pulse trains on the same photodiode leads to the formation of three RF combs (RBW 3~kHz).
\textbf{(e)} Envisioned application of the triple-comb generator for two-dimensional spectroscopy~\cite{Lomsadze2018}
}
\label{F6}
\end{figure*}
% =============================================

%%%%%%%%%%%%%%%%%%%%%%%%%%%%%%
\section*{Introduction}
%%%%%%%%%%%%%%%%%%%%%%%%%%%%%%
Shortly after the inception of the optical frequency comb\cite{Hansch2006a}, it was realised that combining two combs with slightly different repetition rates on a photodetector produces a radio-frequency (RF) interferogram that samples the optical response\cite{Keilmann2004,Coddington2016}, without any moving parts.
Such dual-comb techniques have been demonstrated in both real-time\cite{Schliesser2005,Ideguchi2012} and mid-infrared\cite{Villares2014} spectroscopy, distance measurements\cite{Coddington2009a}, two-way time transfer\cite{Sinclair2017}, coherent anti-Stokes Raman spectro-imaging\cite{Ideguchi2013}, as well as photonic analogue to digital conversion\cite{Ataie2015}.
However, facing the complexity and cost associated with operating two laser frequency combs, novel methods are being actively explored with a view to reduce the system complexity and inherently improve the mutual coherence. For example, instead of phase locking two independent conventional mode-locked lasers, both combs can be generated in the same laser cavity\cite{Link2017}, via repetition rate switching of a single comb\cite{Carlson2018}, or spectrally broadened in the same fibre in opposite propagation directions\cite{Millot2015}. As the noise sources are common mode, the relative coherence between the combs is significantly improved, allowing for longer coherent averaging\cite{Coddington2016}.

Recent advances in the field of high-quality-factor microresonators pumped with a continuous wave (CW) laser have led to the discovery of `Kerr' frequency combs\cite{Del'Haye2007} (also termed `microcombs') that arise due to nonlinear wave mixing mediated by the optical Kerr effect. One particular state of such combs corresponds to the formation of dissipative Kerr solitons\cite{Herr2013} (DKSs) --- self-localised pulses of light circulating in the resonators arising from the double balance between loss and parametric gain and between dispersion and nonlinearity\cite{Lugiato1987,Leo2010}.
The coherent and broadband properties of DKS-based microcombs have already found applications in parallel coherent communication\cite{Marin-Palomo2017}, distance measurement\cite{Trocha2018,Suh2018}, astrophysical spectrometer calibration\cite{Suh2018,Obrzud2017a}, self-referencing\cite{Jost15, Brasch2017}, and photonic-integrated frequency synthesis\cite{Spencer2018}.
DKSs display rich nonlinear dynamics, such as dispersive wave emission\cite{Brasch2015}, Raman self-shifting\cite{Skryabin2003,Karpov2016a}, or breathing solitons\cite{Lucas2017b}.

The initial demonstrations of dual-microcomb applications relied on pairs of physically distinct yet almost identical resonators\cite{Dutt2018,Suh2016,Trocha2018,Marin-Palomo2017,Pavlov2017}.
Recent works\cite{Joshi2018,Yang2017} demonstrated the generation of dual-DKS combs with counter-propagating solitons within the same spatial mode of a single microresonator, using the clockwise and counter-clockwise mode degeneracy, and showed a drastic improvement of the coherence.
However, this technique is limited to counter-propagating pumps and as such requires nonreciprocal elements, i.e. circulators. Moreover, since the same mode family is used, only small relative combs offsets are possible, while the repetition rate difference is induced via the Kerr and Raman effects\cite{Karpov2016a} and remains relatively moderate. As a result, the corresponding RF comb is not centred at sufficiently high frequencies, and RF lines near DC may overlap, which can lead to soliton-locking\cite{Yang2017}, but also implies that several pairs of lines beat at identical RF frequencies. Likewise, the small repetition rate difference restricts the acquisition speed. Finally, the scheme is inherently limited by the twofold degeneracy of whispering gallery modes (WGM), only allowing dual-comb generation. 

Like optical fibres, optical microresonators can also exhibit multiple spatial mode families, which provide additional degrees of freedom in which light can propagate. In fibre optical communication, space-division multiplexing utilizes different spatial modes of an optical fiber as additional parallel channels to transmit data\cite{Richardson2013,Bozinovic2013}. It remains an open question if dual-comb can be generated in an analogous way within different spatial modes of a cavity.
The presence of higher-order modes is known to impact DKSs, causing alterations of the comb envelope via avoided mode crossing\cite{Herr2013b}, dispersive wave emission\cite{Matsko2016a,Yang2016c}, repetition rate instability\cite{Lucas2017,Lucas2017d}, intermode breathing\cite{Guo2017d}, and Stokes soliton generation\cite{Yang2016b}.  Although pumping of two orthogonally polarised modes was investigated in preliminary works\cite{Bao2017a,Donvalkar2017,Zhao2017}, generating independent soliton states in distinct spatial modes was not shown to date.

Here, we demonstrate spatial multiplexing of DKS states in a microresonator. 
We simultaneously generate multiple DKS-based combs within distinct spatial modes of a single optical microresonator. Up to three different mode families of the same polarisation are pumped using a laser and modulation sidebands (\fref{F1}c).
This spatial multiplexing thereby allows not only dual but also triple frequency comb generation from one and the same device, which to the best of the authors knowledge, has not been achieved with any other laser frequency comb platform to date (e.g. Ti:Sa, semiconductor or fibre-based mode locked lasers).
The technique introduced here also overcomes several of the previous shortcomings.
The distinct free spectral ranges of the respective mode families enable the generation of independent soliton pulse streams with substantial repetition rate differences (100~kHz -- 100~MHz). As a single laser and resonator are used, the resulting combs have thus excellent mutual coherence, and support dual-comb spectroscopy (with amplitude and phase retrieval) in spite of using a free running system. The larger offset between combs prevents soliton locking\cite{Yang2017} and associated mapping ambiguity. The multiplexing can also be performed in co- or counter-propagating directions. Finally, the scalability is demonstrated with the generation of three combs in a single resonator. Beyond established dual-comb techniques\cite{Keilmann2004,Coddington2016}, triple comb can be used for higher dimensional spectroscopy, with the potential to increase information content, accuracy or speed of acquisition, such as in 2D coherent spectroscopy\cite{Lomsadze2018,Cundiff2013}, as well as advanced comb-based distance measurement with increased ambiguity range\cite{Zhao2018}. The multiplexing approach could lead to significant simplifications in the implementation of these schemes.

%%%%%%%%%%%%%%%%%%%%%%%%%%%%%%
%%%%%%%%%%%%%%%%%%%%%%%%%%%%%%
\section*{Results}
%%%%%%%%%%%%%%%%%%%%%%%%%%%%%%
%%%%%%%%%%%%%%%%%%%%%%%%%%%%%%
We take advantage of the multi-mode nature of a crystalline \ce{MgF2} WGM cavity (\fref{F1}a). 
The fabrication of \ce{MgF2} crystalline cavities by diamond turning and subsequent polishing with diamond slurries leads typically to multimode resonators with several mode families reaching ultrahigh quality factors ($Q$) exceeding $10^9$. 
In the present work, the resonator used has a free spectral range (FSR) of 12.4~GHz, and features up to 5 mode families with the same polarisation that sustain DKS formation ($Q \geq 10^9$), as shown in \fref{F2} of the supplementary information (SI). The light is coupled using a tapered optical fibre, whose position is adjusted to tune the relative coupling rate to higher order modes.

%%%%%%%%%%%%%%%%%%%%%%%%%%%%%%
\noindent\textbf{Spatial multiplexing with co-propagating pump fields}  ---
%%%%%%%%%%%%%%%%%%%%%%%%%%%%%%
We first study the use of co-propagating pumps. In this scheme (\fref{F1}d,e), simultaneous pumping of two soliton-supporting resonances is achieved via electro-optical modulation. The light of an initial pump laser (external cavity diode laser, wavelength 1554~nm) passes through an IQ-modulator to generate a single sideband~\cite{Izutsu1981}, without fully suppressing the carrier, such that both reach the same power level. This creates two mutually phase-coherent carriers with a tunable frequency offset. The modulation frequency is set to $f_m = \omega_m/2\pi \sim 4.28$~GHz to match the separation of two soliton-supporting resonances, which belong to different spatial mode families but have the same polarisation. The `laser scanning technique'~\cite{Herr2013} is subsequently applied on the main pump laser to trigger DKS formation in both mode families simultaneously. A successful tuning is however challenging as each resonance induces a thermal shift. This is mitigated in two ways: the laser is tuned across the resonances using the diode current, which allows tuning speeds faster than the thermal relaxation time of the cavity (typically ms timescale). Second, the modulation frequency $f_m$ is carefully adjusted so that the `high detuning end' of both soliton steps~\cite{Lucas2017} approximately coincide (\fref{F2}b-c). We observed that this maximises the success rate of simultaneous single DKS initiation in both mode families (see SI for details).
After generation, the main laser is locked to the microresonator via offset Pound-Drever-Hall (PDH) locking~\cite{Lucas2017d} and the dual-DKS combs can be stably maintained for more than 12 hours.

\newcommand{\powten}[1]{\times 10^{#1}}
In this manner two simultaneous streams of DKSs are produced. The optical spectrum of the microresonator output (\fref{F3}a) shows the two interleaved DKS combs offset by $f_m$. The repetition rates of the two combs differ by $\DFRep = 655$~kHz (around 12.4~GHz). This corresponds to a relatively small spectral compression factor~\cite{Coddington2016} of $m = f_{\rm rep}/\DFRep = 1.8 \powten{4}$, which is useful to increase the acquisition speed of a moderate optical span.
The beating of the dual-comb results here in an RF comb centred at $f_m = 4.28$~GHz. The individual lines of the RF comb are still resolution-limited at 100~Hz bandwidth, although the system is free running (neither the combs' repetition rates nor the pump wavelength are stabilised). In comparison, two combs generated in distinct microresonators pumped with the same laser have a linewidth broader than 1~kHz. Thanks to the high centre frequency $f_m$, the total $\sim 200$~MHz span of the RF comb can be mapped into the corresponding 3~THz of optical span without aliasing at baseband frequencies and no signs of inter-soliton locking~\cite{Yang2017} were observed consequently. Nevertheless, it may seem that a high centre frequency is a drawback, as it requires the use of a very fast sampling rate to directly acquire the dual comb interferogram, but since this frequency is set by the modulation $f_m$, it is possible to downmix the RF comb to baseband with an IQ demodulator and relax the requirement on the sampling rate.

Importantly, although the solitons circulate in distinct spatial modes, they can interact via four-wave mixing (FWM) when co-propagating and effectively become modulated at the rate at which the solitons cross $\DFRep$, as detailed in the supplementary information (SI, cf. \fref{F8}).  The modulation products that arise around the comb lines will beat with adjacent comb lines at frequencies identical to the RF comb and may thus induce optical-to-RF mapping ambiguities. In the present case, the relative strength of the first intermodulation sideband is approximately $ -20$~dBc (see \fref{F8}e), in agreement with the effect of cavity filtering. This value is sufficiently weak to be neglected in most practical applications.

Another pair of mode families can be selected to achieve a larger repetition rate difference (see \fref{F4}, $\DFRep = 9.3$~MHz) when a faster acquisition is targeted. The compression $m = 1.4\powten{3}$ is more than one order of magnitude lower than the previous demonstrations with counter-propagating solitons ($3.3\powten{5}$ in~\cite{Yang2017} and $3.2\powten{4}$ in~\cite{Joshi2018}), whilst typical mode-locked lasers are in the range~\cite{Coddington2016} $3\powten{4} - 10^{6}$.
Note that increasing $\DFRep$ enables even stronger intermodulation suppression ($\sim -40$~dBc here) as the sidebands are created well outside the cavity bandwidth (see SI, \fref{F8}f).
These experiments illustrate the flexibility of the technique and its potential to substantially increase the bandwidth of the dual-comb interferogram and acquisition speed, compared with prior schemes using counter-propagating solitons~\cite{Yang2017}.
Although a collinear dual-comb is not suitable for some applications, co-propagating soliton generation simplifies the scheme considerably, as it lifts the need for nonreciprocal devices. Furthermore,  two combs generated this way could be separated via de-multiplexing, if the offset $f_m$ is high enough (e.g. beyond 25~GHz), which should be possible for integrated micro-resonators with larger FSR ($> 100$~GHz). Alternatively, if the pumped modes are orthogonally polarized, they could be demultiplexed with a simple polarisation beam splitter.

%%%%%%%%%%%%%%%%%%%%%%%%%%%%%%
\noindent\textbf{Spatial multiplexing in the counter-propagating configuration} ---
%%%%%%%%%%%%%%%%%%%%%%%%%%%%%%
Alternatively, the two spatial mode families can be excited in a counter-propagating way, analogous to previous implementations~\cite{Joshi2018,Yang2017}. First, the pump laser is split unevenly between two paths. A pair of circulators is then used to couple light into the resonator and to collect the transmitted combs on both sides (\fref{F5}). 90\% of the amplified pump power ($\sim 200$~mW) is coupled directly into the counter-clockwise (CCW) direction. In the other path, the remaining 10\% of the pump is sent through a single sideband modulator operated in carrier-suppressing mode to frequency-shift the light by the offset separating the two resonances. After amplification to a similar power level of $\sim 200$~mW, the frequency-shifted light is coupled in the clockwise (CW) direction.

We used another set of mode families whose resonance offset is $f_m = 2.75$~GHz, and the repetition rate difference is $\DFRep = 371$~kHz to demonstrate the counter-clockwise spatial multiplexing of DKSs. The soliton formation is triggered in the same way as in the co-propagating case. The two generated single-soliton combs are shown in \fref{F5}. Each comb has an average power of $\sim 400~\upmu$W at the output of the resonator (after excluding the pump). The corresponding RF comb (\fref{F5}) features a similar degree of stability to the co-propagating scheme, with 200~Hz wide beatnotes throughout the RF comb (\fref{F5}d).
A main advantage of the counter-propagating pump configuration, is the absence of intermodulation products in the combs (\fref{F8}g). This is to be expected, as in this case, the  FWM process between two different combs is momentum-forbidden (see SI for details).

With this pumping configuration the combs can also be accessed individually, allowing the implementation of a wider range of dual-comb applications.
As a proof of concept spectroscopy experiment, the combs are amplified to $\sim 10$~mW average power and one comb is sent through a waveshaper before interfering with the second comb. The beating is recorded on a high sampling rate oscilloscope (1 ms acquisition time, corresponding to $\sim 370$ averages). The amplitude and phase of the RF comb teeth are compared to a reference signal recorded without the waveshaper. \Fref{F5} shows that the retrieved amplitude and phase closely match the programmed synthetic resonance profiles over a span of 4~THz.

Rapid coherent linear optical sampling~\cite{Coddington2009} was also realised to resolve the dynamics of a DKS pulse breathing~\cite{Bao2016a,Lucas2017b} at a rate $\sim 1$~MHz. Indeed, the fast recording of the interferogram between a DKS comb and a reference comb offers the possibility to spectrally resolve the soliton dynamics in the microresonator, as recently demonstrated~\cite{Yi2018} using an electro-optic comb as reference.
Here, the multiplexing scheme in counter-propagation is applied instead and the detuning in each direction is carefully set in order to generate a CCW breathing pulse and a stable DKS in the CW direction that serves as a reference (\fref{F10}a). Another pair of mode families yielding a higher repetition rate difference ($\DFRep = 9.3$~MHz) is selected, such that the acquisition speed of the interferogram is faster than the breathing rate. The optical spectra of each pulse train can be viewed in panels b and c. The breathing soliton spectrum features a typical triangular profile on the optical spectrum analyser due to the averaging of the periodic spectral broadening and compression. The real-time spectral evolution of the breathing soliton can be retrieved by taking the Fourier transform to the dual comb interferogram (\fref{F10}d, see Methods for details). The salient features of breathing DKS can be retrieved~\cite{Bao2016a}: over half a breathing period, the spectrum contracts and expands. Furthermore, the comb lines located near the pump are oscillating out of phase from the wing.
The detection of the dual comb interferogram envelope allows the spatiotemporal dynamics of the breathing soliton to be mapped. After accounting for the compression ratio of the dual comb acquisition, this methods yields an effective temporal resolution of $\sim 1$~ps (\fref{F10}f) and represents a 10 fold improvement compared to the direct real-time sampling method (\fref{F10}e, see SI for details).

%%%%%%%%%%%%%%%%%%%%%%%%%%%%%%
\noindent\textbf{Triple soliton comb generation via spatial multiplexing} ---
%%%%%%%%%%%%%%%%%%%%%%%%%%%%%%
We next demonstrate the ability to multiplex three soliton combs by pumping three mode families simultaneously.
We employ the counter-propagating configuration, but combine two tones on the modulator (\fref{F5}): $f_m = \omega_m/2\pi = 2.17$~GHz and $f_m' = \omega_m'/2\pi = 4.59$~GHz. This allows two mode families to be co-pumped, and thus the creation of two combs in the CW direction, while another comb is generated by pumping a third mode family in the CCW direction.
Remarkably, the excitation technique outlined earlier, was also applied successfully to generate all three single soliton state combs (\fref{F6}a,b).
Heterodyning the combs creates a set of three RF combs centred at $f_m$, $f_m'$ and $|f_m'-f_m| = 2.42$~GHz and with a line spacing of 380~kHz, 355~kHz, and 735~kHz respectively.
In the heterodyne RF combs between the CCW and each of the CW combs, we are able to observe weak additional lines resulting from the intermodulation products on each of the CW soliton combs (\fref{F6}b,d). The spikes around $\pm 10$~MHz on these RF-combs are caused by the PDH phase modulation. Importantly, these additional beatnote products give rise to spectrally distinct frequencies and can thus be removed during signal processing (by only selecting the frequency components at e.g. $f_m + n \, \Delta f_{\rm rep}$), and critically do not induce mapping ambiguities.

The triple soliton comb configuration with two co-propagating combs could find applications in advanced spectroscopy schemes such as two-dimensional spectroscopy~\cite{Cundiff2013}. A recent demonstration~\cite{Lomsadze2017,Lomsadze2018} employed three Ti:Sa lasers, two of which generated a pump and probe pulse trains to excite a photon echo in Rubidium vapour, which was heterodyned with the third local oscillator comb allowing the fast acquisition of 2D spectra with a single photodiode. The multiplexing approach would offer a major simplification and cost reduction of such schemes, as illustrated in \fref{F5}e. Optical distance measurements can also benefit from this triple comb scheme, as sending a pair of combs onto the target would provide two series of synthetic wavelength chains, allowing a great extension of the ambiguity range~\cite{Coddington2009a} without compromising the resolution. Such scheme was recently demonstrated using three electro-optic combs~\cite{Zhao2018} reaching an accuracy of 750~nm over 80~m distance (instead of 15~mm with the dual comb method).

%%%%%%%%%%%%%%%%%%%%%%%%%%%%%%
\section*{Discussion}
%%%%%%%%%%%%%%%%%%%%%%%%%%%%%%
In summary, spatial multiplexing of soliton combs in a single microresonator is demonstrated experimentally for the first time, both in co- and counter-propagating pump configuration.
The multiple soliton pulse streams have excellent mutual coherence, and their frequency offset is a substantial fraction of the repetition rate, which prevents mapping to negative frequencies. The generated dual-combs are shown to be suitable for spectroscopy.
Large relative differences in repetition rates can be obtained, enabling fast acquisition and improved bandwidth usage. Using the birefringence  of \ce{MgF2} even larger differences are possible, when pumping modes with orthogonal polarisations ($> 110$~MHz were observed, as demonstrated in the SI). When combined with faster repetition rates, such configurations could find applications in RF signal processing and acquisition~\cite{Esman2016}, or for ultra-rapid vibrational spectroscopy in condensed matter.
The fast recording of a dual DKS-comb heterodyne also prove useful to investigate soliton dynamics with unprecedented resolution, such as measuring the line-by-line spectral dynamics of a breathing soliton.
We also demonstrate that a high repetition rate difference is also beneficial to suppress the intermodulation products when the solitons are co-propagating (as shown in \fref{F8}c).

The presented scheme is already within reach of microfabricated ring waveguide resonators~\cite{Kim2017,Pfeiffer2017}, as illustrated by the recent demonstration of a device supporting solitons in both the TE00 and TM00 mode families~\cite{Liu2018}, for two closely spaced pump frequencies. In the future, this improved fabrication control will allow the control of the mode frequency separation and repetition rate difference, while mitigating the impact of modal crossing.
Furthermore, waveguide geometric dispersion engineering will enable larger bandwidth coverage~\cite{Pfeiffer2017,Brasch2015} and central wavelength selectivity~\cite{Lee2017,Karpov2018a}.
The simplicity of the co-propagating scheme makes it compatible with full on-chip integration, as all the elements are readily available in photonic integrated circuits.

The method is flexible and easily scalable, as shown by the generation of three simultaneous soliton combs -- so far out of reach for other frequency comb platforms.
This multiple comb source has the potential to extend the capabilities of comb-based method, for greater information content, accuracy or speed of acquisition.

\section*{Methods}
\noindent\textbf{Resonator fabrication and characteristics} ---
The microresonator protrusion was fabricated via high-precision diamond turning of a mono-crystalline \ce{MgF2} blank followed by hand polishing with diamond slurries and cleaning. The FSR of 12.4~GHz corresponds to a major radius of 2.8~mm. The resulting WGM protrusion can be approximated by an oblate spheroid with a width of $25~\mathrm{\upmu m} $ and a height of $ 80~\mathrm{\upmu m} $ (this very shallow protrusion appears almost flat in the picture \fref{F1}a). While single mode cavity protrusions have been demonstrated~\cite{Grudinin2015} using advanced micro-machining techniques, we target a wide protrusion instead, which supports a large number of WGM modes and makes the polishing less challenging. The obtained quality factors of the soliton-supporting resonances are above $10^9$ at 1554~nm.
The group velocity dispersion (GVD) of \ce{MgF2} is naturally anomalous in the C-band ($\beta_2 \sim -9.1~\rm fs^2/mm$), so that no geometric dispersion engineering is needed to reach this dispersion regime, which is necessary for soliton formation. Due to the loose confinement and the large main radius of the structure, higher order modes have essentially a higher FSR without significantly changing the dispersion which is dominated by the material. Overall, 5 resonators were fabricated, featuring at least two soliton-supporting mode families (FSR 8, 12, 14, 17 and 26~GHz).

Evanescent coupling to the WGM is achieved with a tapered optical fibre. The mode of the tapered fibre and of the WGM resonator are not orthogonal to each other and one taper mode can excite several higher order WGM, although the mode overlap and phase matching condition are different for each WGM thus leading to variations of their respective coupling rate. However, shifting the optical fibre position out of the equatorial plane of the resonator, influences the coupling strength of the even and odd WGMs~\cite[p.~59]{2005Humphrey} (in the polar direction). This degree of freedom is used in the experiment to adjust the coupled power in the modes producing the solitons.

\smallskip\noindent\textbf{Dual comb imaging of the soliton dynamics} ---
In a specific range of detuning and pump parameters, DKS can undergo a  breathing instability where the solitonic pulse oscillates both in amplitude and duration~\cite{Bao2016a,Lucas2017b}. Thanks to the long photon lifetime of microresonators, the oscillation period is much longer than the roundtrip time (the breathing rate is $\sim 1$~MHz, for a 12.7~GHz repetition rate).

First, resolving the spatiotemporal dynamics of breathing solitons was carried out via direct sampling of the breathing pulse train, using a very fast photodiode ($\sim 10$~ps impulse response) and real-time oscilloscope (120~GSa/s) in order to sample every roundtrip~\cite{Lucas2017b}. However, the fast temporal resolution of this method is limited by the photodiode impulse response and closely spaced solitons may not be distinguished (\fref{F10}e), while the slow breathing evolution is oversampled and can only be monitored for a short time span at such sampling rates.

Another method is to take advantage of the dual comb principle. This method is derived from coherent linear optical sampling~\cite{Duran2015,Coddington2009}. The envelope of a dual comb interferogram between a breathing soliton and a reference pulse train with a slight difference in repetition rate $\DFRep$ yields the convolution of the breathing pulse with the reference. 
The pulses overlap each $1/\DFRep$, which corresponds to the time for one pulse to sweep over one entire roundtrip of the other pulse train, and sets the imaging frame rate.
Thus if $\DFRep$ is faster than the breathing rate, the breathing dynamics can be monitored, while relaxing the requirement on the sampling rate (ultimately only to match RF comb bandwidth).

The dual comb is generated via multiplexing two solitons in counter-propagation (see \fref{F10}). The mode families are selected to reach $\DFRep = 9.26$~MHz (one frame period is acquired over 1340 roundtrips). The real-time spectral evolution of the breathing soliton (multiplied by the reference pulse spectrum) can be retrieved by taking the Fourier transform of the dual comb interferogram (\fref{F10}d). In order to improve the signal to noise ratio, multiple interferogram frames at similar breathing phases were averaged together (after multiplication by a gaussian window of width $1/\DFRep$).

To view the spatiotemporal dynamics of the breathing soliton, the interferogram envelope is retrieved via Hilbert transform and each frame is sliced and stacked (\fref{F10}f). The fast time axis can be rescaled to span $1/f_{\rm rep}$ to account for the compression ratio of the dual comb acquisition method. Panels e and f show the vast improvement in temporal resolution of the dual comb method over the direct sampling method ($\sim 1$~ps vs. $\sim 10$~ps).

%%%%%%%%%%%%%%%%%%%%%%%%%%%%%%%%%%%%%%%%%%%%%%%%%%%%%%%%%%%%%%%%%%
%%%%%%%%%%%%%%%%%%%%%%%%%%%%%%%%%%%%%%%%%%%%%%%%%%%%%%%%%%%%%%%%%%
\section*{Data availability statement}
The data and code used to produce the results of this manuscript will be available on Zenodo upon publication.

%%%%%%%%%%%%%%%%%%%%%%%%%%%%%%%%%%%%%%%%%%%%%%%%%%%%%%%%%%%%%%%%%%
%%%%%%%%%%%%%%%%%%%%%%%%%%%%%%%%%%%%%%%%%%%%%%%%%%%%%%%%%%%%%%%%%%
\section*{Authors contributions}
E.L. and G.L. designed the experimental setup. E.L. performed the experiments and analysed the data. G.L. fabricated the device, with assistance of N.G.P.. E.L., R.B. and A.R. performed the experimental comb linewidth measurement.
M.K. and A.R. assembled the RF components for the single sideband modulator driving.
E.L. wrote the manuscript, with input from other authors. T.J.K. and M.L.G. supervised the project.

%%%%%%%%%%%%%%%%%%%%%%%%%%%%%%%%%%%%%%%%%%%%%%%%%%%%%%%%%%%%%%%%%%
%%%%%%%%%%%%%%%%%%%%%%%%%%%%%%%%%%%%%%%%%%%%%%%%%%%%%%%%%%%%%%%%%%
\begin{acknowledgments}
The authors thank Dr. Nathan Newbury for important suggestions and comments. The authors thank J.D. Jost and W. Weng for their assistance as well as J. Liu, H. Guo, N.J. Engelsen and M. Anderson for their feedback on the manuscript.
This publication was supported by funding from the Swiss National Science Foundation under grant agreement 163864, by the Air Force Office of Scientific Research, Air Force Material Command, USAF under Award No. FA9550-15-1-0099, and by the Ministry of Education and Science of the Russian Federation under project RFMEFI58516X0005. E.L. acknowledges the support of the European Space Technology Centre with ESA Contract No.: 4000118777/16/NL/GM.
\end{acknowledgments}

%%%%%%%%%%%%%%%%%%%%%%%%%%%%%%%%%%%%%%%%%%%%%%%%%%%%%%%%%%%%%%%%%%
%%%%%%%%%%%%%%%%%%%%%%%%%%%%%%%%%%%%%%%%%%%%%%%%%%%%%%%%%%%%%%%%%%
% Bibliography
%\bibliography{Biblio}
\putbib

\end{bibunit}

\clearpage

\begin{bibunit}

%%%%%%%%%%%%%%%%%%%%%%%%%%%%%%%%%%%%%%%%%%%%%%%%%%%%%%%%%%%%%%%%%%
%%%%%%%%%%%%%%%%%%%%%%%%%%%%%%%%%%%%%%%%%%%%%%%%%%%%%%%%%%%%%%%%%%
\section*{Supplementary Information}
\renewcommand\thefigure{S.\arabic{figure}}    
\setcounter{figure}{0}

\begin{figure}[h]
\centering
\includegraphics[width=\columnwidth]{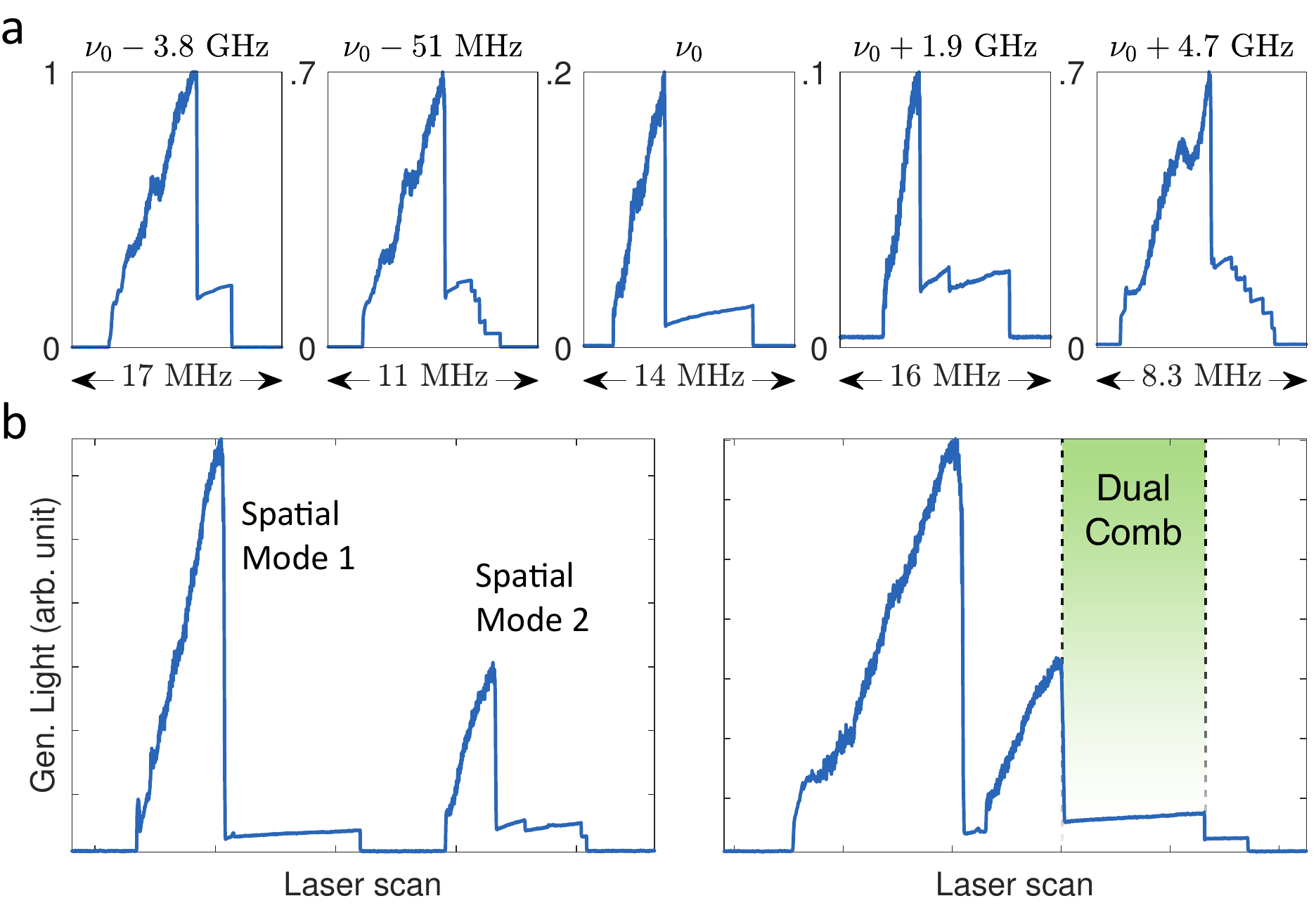}
\caption{
\textbf{(a)}~Identification of the soliton-supporting resonance over one cavity FSR. The graphs display the generated comb light at the output of the resonator as the laser frequency is decreased. The step features correspond to the detuning region where solitons exist.
\textbf{(b)} Sequential excitation of two soliton-supporting resonances (in the co-propagating scheme), when the offset frequency is detuned. The left resonance is excited with the pump laser light while the right is excited with the sideband.
\textbf{(c)} Adjusting the sideband shift allows the overlap of the resonances. The region where the two steps coexist corresponds to the formation of the dual DKS comb.
}
\label{F2}
\end{figure}

\smallskip\noindent\textbf{Identification of soliton resonances} ---
The piezo of the pump ECDL is scanned over a full FSR of the cavity at 1554~nm, while recording the transmission and the nonlinearly generated light on an analogue to digital converter.
%Around 100 resonances were recorded and 
Five resonances featuring the typical step transition associated with the formation of solitons~\cite{Herr2013} could be identified (\fref{F2}a). Their relative offset was estimated using piezo voltage calibration.

\smallskip\noindent\textbf{Tuning Method} ---
In order to initiate the dual-comb formation, a pair of target mode families is first identified. The pump laser is tuned close to one resonance and the modulation frequency of the sideband is set close to the resonance offset, while the bias of the SSB modulator is adjusted to generate a blue or red sideband, depending on the sign of the frequency shift needed. After a coarse adjustment, both resonances are visible when scanning the pump laser over a small frequency span (\fref{F2}b). The coupling can be optimized by changing the tapered fibre position in order to increase the soliton step length in both families. The final adjustment consists of tuning the sideband frequency shift so that the large detuning end of each the soliton step becomes aligned (\fref{F2}c). When scanning the laser, two soliton states can then be excited simultaneously more than 50 \% of the time. Upon tuning the laser across resonance, the number of generated solitons in one state is typically stochastic, due to the chaotic modulation instability that seeds the solitons~\cite{Herr2013}. However, the single soliton state is the most attractive state due to its smooth envelope, but remains challenging to obtain in microresonators. We found that our procedure also improves the success rate for dual single soliton generation. We believe this is because the tuning to the ‘high detuning end’ makes it more favourable for solitons to collide or decay~\cite{Zhou2015,Luo2015}. Nonetheless the dual single soliton production remains less probable, with an estimated success rate below 10 \%. This rate could be improved by implementing an active capture feedback mechanism\cite{Yi2016}.

% =============================================
% =============================================

% =============================================
\begin{figure*}[t]
\centering
\includegraphics[width=\textwidth]{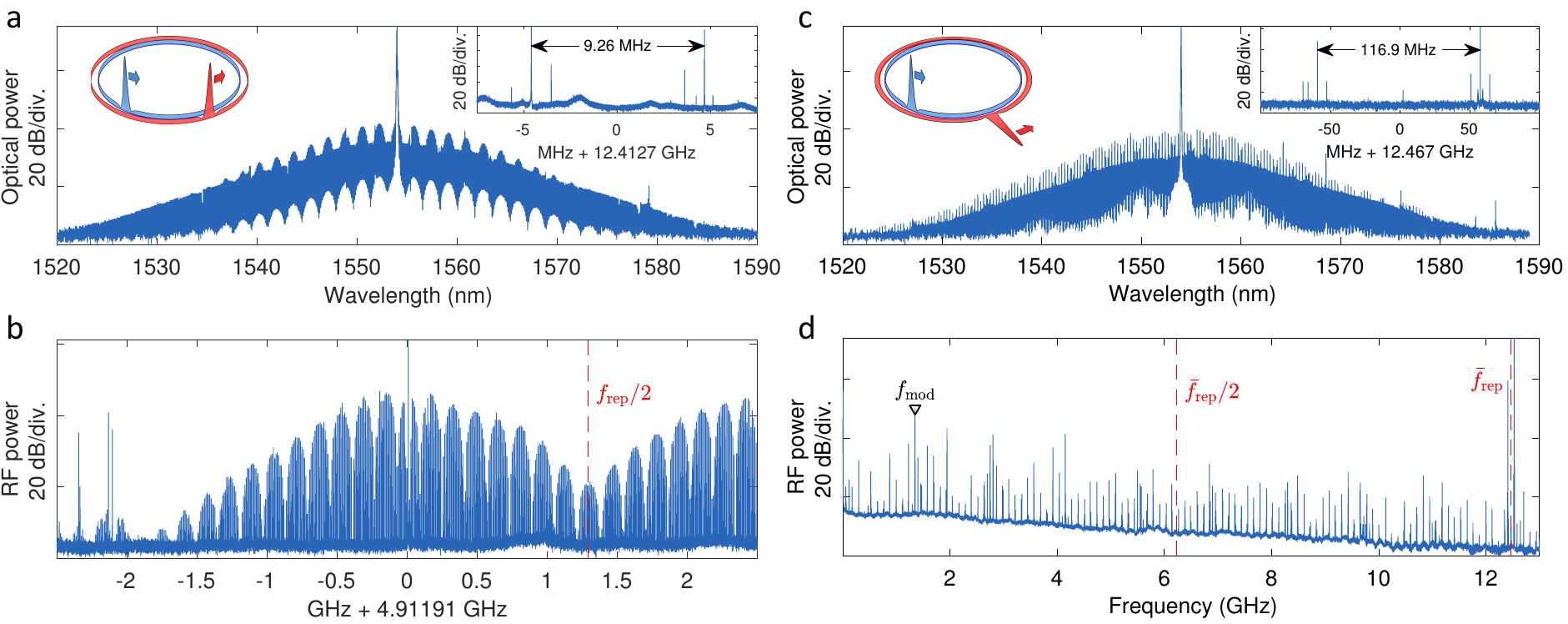}
\caption{
\textbf{(a)} Generated dual-comb optical spectrum in the co-propagating direction. One of the combs corresponds to a two-soliton state and hence has a distinct spectral interference pattern. The repetition rate difference is shown in the inset.
\textbf{(b)}~Corresponding RF heterodyne comb.
\textbf{(c)} Generated dual-comb optical spectrum in co-propagation, for mode families with orthogonal polarization. The repetition rate difference is shown in the inset.
\textbf{(d)}~Corresponding RF heterodyne comb. The modulation frequency is indicated. The RF comb is heavily aliased due to the insufficient bandwidth for such large repetition rate difference.
}\label{F4}
\end{figure*}
% =============================================

\smallskip\noindent\textbf{Larger repetition rate difference via selection of mode families} ---
Several RF comb offset frequencies and repetition rate differences can be achieved in the same resonator, by changing the pair of modes supporting the solitons. In this way we could generate solitons in the co-propagating direction (\fref{F4}a) with an offset of 4.9~GHz and a repetition rate difference of $\sim 9$~MHz. The resulting RF comb (\fref{F4}b) spans more than 4~GHz. However, this high offset frequency $f_m$ combined with a broader comb implies that the RF comb extends beyond $f_{\rm rep}/2$ and thus overlaps with the mirror comb~\cite{Coddington2016} centred at $f_{\rm rep} - f_m$, leading to potential mapping ambiguities in the overlap region. Engineering the modes of the microresonator, enabled by better fabrication control, will allow an optimal bandwidth usage.

\smallskip\noindent\textbf{Very large repetition rate difference via pumping of orthogonally polarised modes} ---
The microresonator not only supports higher order spatial modes but also fundamentally orthogonally polarized modes. Furthermore, as \ce{MgF2} is birefringent ($n_o \sim 1.37$ and $n_e \sim 1.38$ at 1554 nm), and the axis of rotation of the WGM resonator is oriented along the optical c-axis, two orthogonally polarized modes feature a greater difference in their free-spectral range. Note that the material group velocity dispersion is anomalous in both direction. We demonstrate the generation of two co-propagating soliton states in orthogonally polarised modes (resonance separation $f_m = 1.34$~GHz). The simultaneous pumping of both modes is achieved by aligning the polarisation of the pumps at $45^\circ$ with respect to the polarisation of each respective mode. Note that half of the energy of each pump is unused in that case. The pumping efficiency could be improved by first splitting the laser light with a polarising beam splitter, modulating one path with the SSB, and combining both paths before coupling to the resonator. In this way, each pump can be aligned with the respective polarisation mode, allowing for more efficient coupling and avoid energy loss.
The resulting combs have a repetition rate difference of $\sim 117$~MHz (\fref{F4}c), which is too high for the available bandwidth (very small compression factor $m = 106$). As a result, the RF comb is heavily aliased at baseband frequency and with the image comb centred around $f_{\rm rep} - f_{m}$, which prevents any application without optical filtering to reduce the bandwidth. Pumping orthogonal mode can also be realised in counter propagation by selecting the proper pump polarisation for each direction.

% =============================================
\begin{figure*}[t]
\centering
\includegraphics[width=\textwidth]{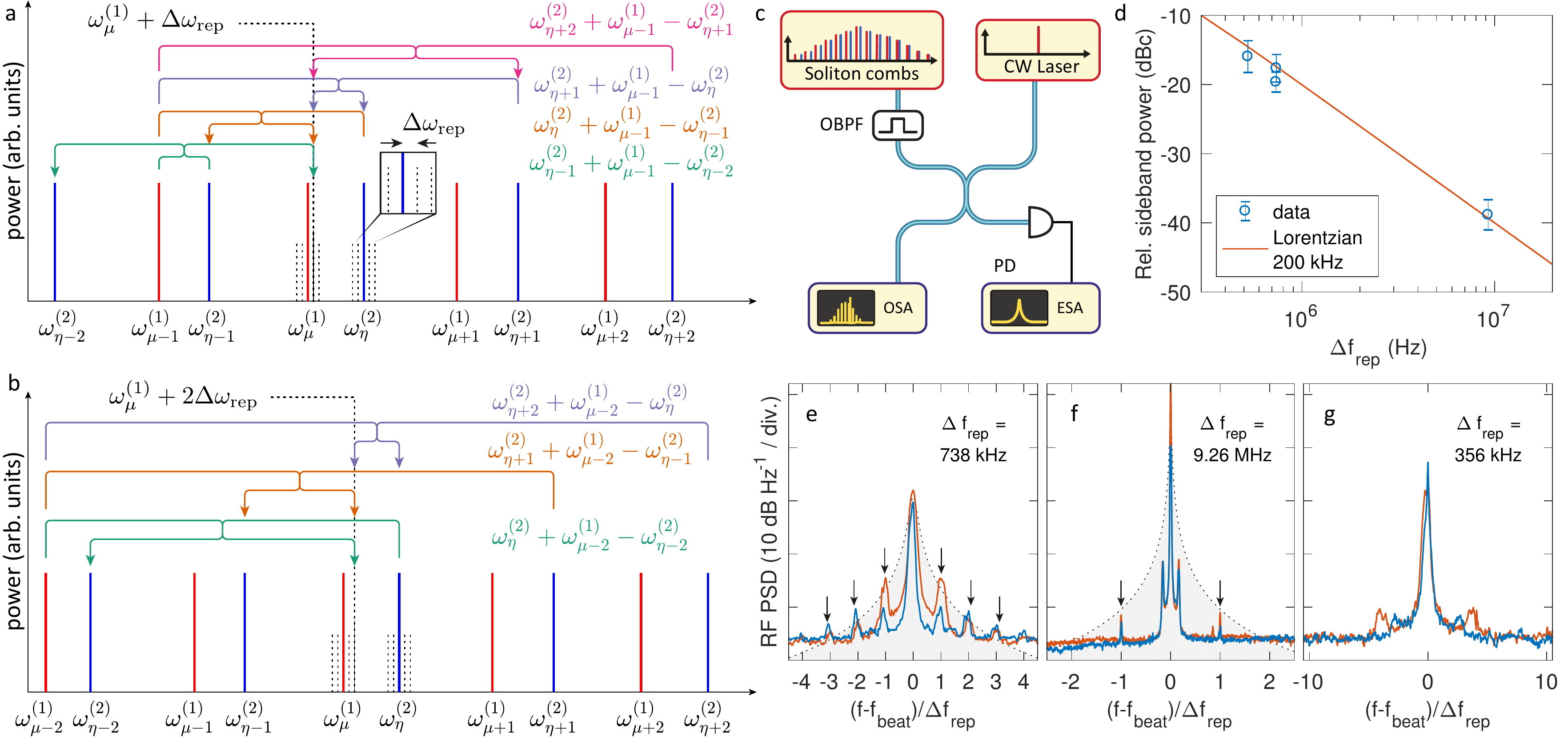}
\caption{
\textbf{Intermodulation of co-propagating solitons}
\textbf{(a)}~Illustration of FWM processes leading to the formation of the $ + \DWRep$ sideband around the comb line $\omega_{\mu}^{(1)}$ (five comb lines are considered here, the cavity filtering is not taken into account).
\textbf{(b)}~Processes leading to the formation of the $ + 2 \DWRep$ sideband around the comb line $\omega_{\mu}^{(1)}$
\textbf{(c)}~Experimental scheme used to measure the optical lineshape of the comb lines. Individual lines of each comb are heterodyned with an independent laser, after selection with an optical bandpass filter (OBPF) and the beat measured on an electronic spectrum analyser (ESA).
\textbf{(d)}~Scaling of the mean relative power in the first sidebands (at $\pm \DWRep$) averaged over several comb lines, for different repetition rate difference $ \DFRep = \DWRep/2\pi$. 
\textbf{(e)}~Heterodyne beatnotes of the reference laser with a line of two co-propagating combs with a repetition rate difference of $\DFRep = 738$~kHz (RBW 50~kHz). The black arrows show the intermodulation sidebands.
\textbf{(f)}~Same measurement in the case of $\DFRep = 9.26$~MHz (RBW 100~kHz).
\textbf{(g)}~Heterodyne beatnotes of the reference laser with a line of two counter-propagating combs with a repetition rate difference of $\DFRep = 356$~kHz (RBW 50~kHz).
}\label{F8}
\end{figure*}
% =============================================

\smallskip\noindent\textbf{Intermodulation products via inter-comb FWM for co-propagating solitons} ---
The two generated dissipative Kerr solitons, exhibiting different free spectral range, can in general interact and cause intermodulation products. Intermodulation of two co-propagating solitons can occur via four-wave mixing, and can lead to additional sidebands around the optical comb lines. We consider here the comb line frequencies in the case of two combs (1) and (2):
\begin{align}
\omega_{\mu}^{(1)} &= \omega_p + \mu \, \omega_{\rm rep}^{(1)} \\
\omega_{\eta}^{(2)} &= (\omega_p + \omega_m) + \eta \, (\omega_{\rm rep}^{(1)} + \DWRep)
\end{align}
where $(\mu, \eta)$ are the azimuthal mode numbers (relative to the pumped mode, for which $\mu = 0$ and $\eta = 0$), $\omega_p$ the pump laser frequency, $\omega_{\rm rep}^{(1)}$ the repetition rate of the first comb, $\omega_m$ the single sideband modulation frequency and $ \DWRep =2\pi \DFRep $ the difference in repetition rate. Inter-comb four-wave mixing can occur for lines fulfilling the phase matching condition (i.e. angular momentum conservation that is $\int d\phi \cdot e^{i(\mu + \eta - \mu' - \eta')\cdot \phi) } = 1$ ): $ \mu + \eta = \mu' + \eta' $.
For counter-propagating solitons in distinct mode families, this momentum matching cannot be satisfied.
However, for two co-propagating  mode families for example,  $ \mu + \eta = (\mu - 1) + (\eta + 1)$ is a possible path that conserves momentum, and the resulting frequencies are $ \omega_{\mu}^{(1)} + \omega_{\eta}^{(2)} = \omega_{\mu-1}^{(1)} + ( \omega_{\eta+1}^{(2)} - \DWRep ) $. As the last frequency does not coincide with an existing comb line, and falls outside the cavity resonance, the mixing product is expected to be inefficient (and suppressed by the cavity lorentzian). Another series of FWM processes leading to the creation of the intermodulation sideband at $\omega_{\mu}^{(1)} + \DWRep$ are represented in \fref{F8}a, when considering 5 comb lines (the cavity filtering is not accounted for).

\begin{figure}[h]
\centering
\includegraphics[width=\columnwidth]{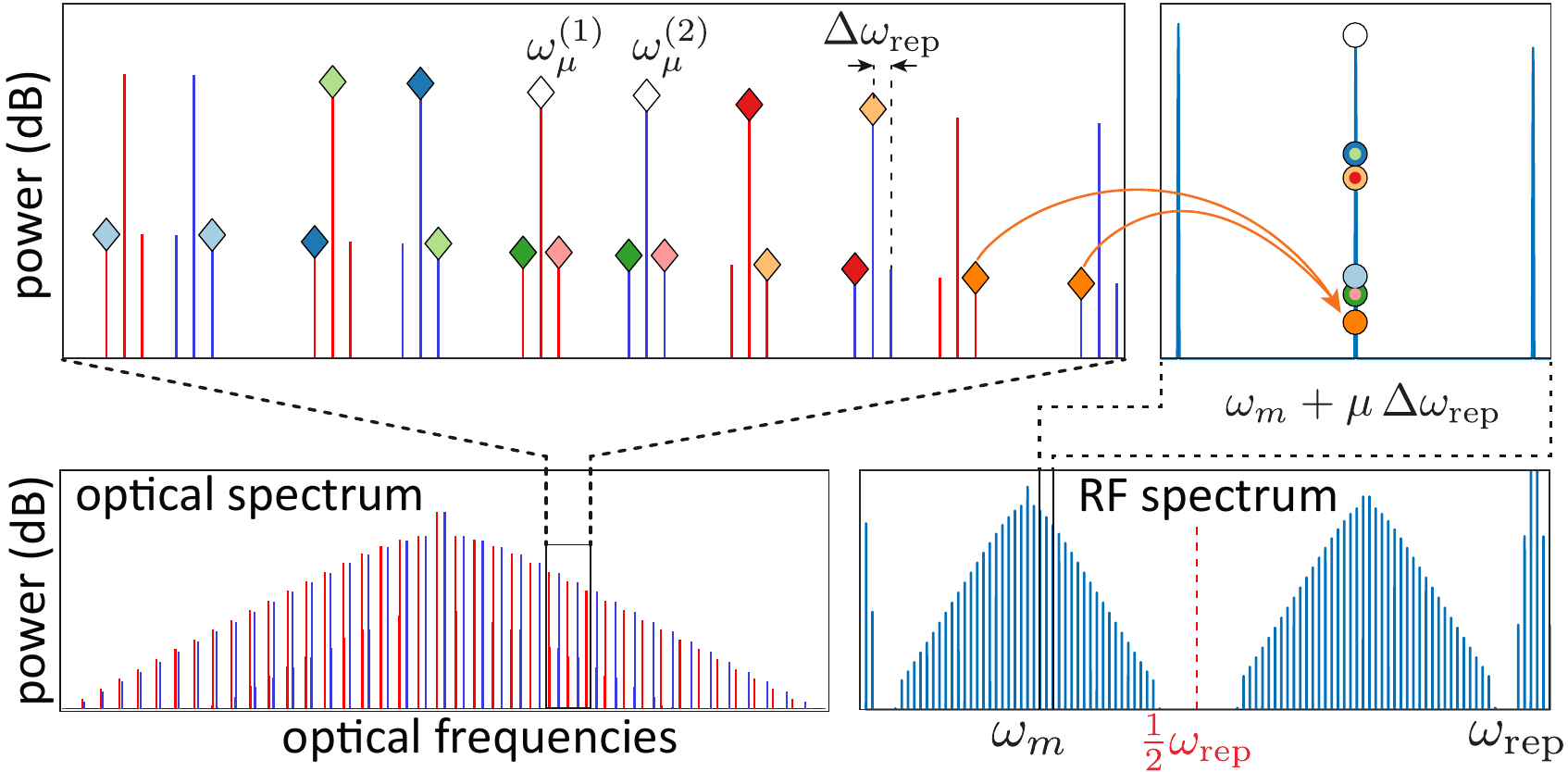}
\caption{
Illustration of the mapping ambiguity induced by the intermodulation sidebands around the comb lines. The lines $ \omega_{\mu}^{(1)} $ and $ \omega_{\mu}^{(2)} $ beat at a frequency $\omega_m + \mu \, \DWRep$, which is the same frequency as the beating between adjacent lines / sidebands. The pairs of optical lines / sidebands beating at the same frequencies are marked with identical colour (top left, the spacing of the sidebands was expanded for visualization) and their corresponding mixing in the RF domain in indicated by a dot with the matching colour (top right).
}\label{F9}
\end{figure}

The presence of sidebands spaced by $\DWRep$ around each of the comb lines can induce optical-to-RF mapping ambiguities, as illustrated in \fref{F9}. The beat between lines  $ \omega_{\mu}^{(1)} $ and $ \omega_{\mu}^{(2)} $ results in the frequency $ \omega_{\mu}^{(2)} - \omega_{\mu}^{(1)} = \omega_m + \mu \, \DWRep$. However, the beat between the adjacent lines $\pm \mu$ and their sidebands as well as a pair of sidebands around $\mu\pm2$ will be at an identical frequency. Therefore, importantly, the presence of sidebands around the comb lines does not appear in the RF dual-comb spectrum, but can be evidenced by the appearance of several lines spaced by $ \DWRep$ around the repetition rates of the combs (or by a high resolution recording of the optical spectrum). Note that in the present experiments the cross products of a comb line and a sideband are attenuated by the relative amplitude of the sidebands i.e. at least 20~dB.

% =============================================
% =============================================

Experimentally, we evaluated the strength of the intermodulation sidebands by beating several lines of each soliton comb with another reference laser centred at 1556.5~nm (\fref{F8}c), having an optical linewidth of $\sim 30$~kHz. If the solitons are co-propagating (\fref{F8}d-f), sidebands at $ \DWRep$ can be clearly identified. This measurement was repeated while pumping a different selection of mode families, such that the scaling of the sideband strength with $ \DWRep$ could be retrieved. The result shown in \fref{F8}d reveal that the mean power of the first sidebands (averaged over multiple comb lines) decreases for larger repetition rate difference, with a slope that matches a lorentzian profile with a typical linewidth of 170~kHz (full width half maximum) which is in line with the measured quality factors of the resonances.

Conversely, when measuring  the optical lineshape of counter-propagating comb lines, no detectable signs of intermodulation products could be observed in any of the $\DWRep$ configurations (\fref{F8}g). In that case, the phase matching condition cannot be fulfilled at the same time as the energy conservation unless $\DWRep = 0$.

\smallskip\noindent\textbf{Stability} ---
In our experiments, the detuning of the laser with respect to one of the pumped modes is actively stabilised via an offset Pound-Drever-Hall (PDH) lock~\cite{Lucas2017d}. This ensures that the resonance-laser  detuning remains within the soliton supporting range~\cite{Herr2013}, as the resonator is free-running and subject to temperature drift. With regards to the relative stability of the produced dual-comb, this means that the main source of instability is the drift of the repetition rates difference $\DFRep$, since the frequency offset between the two pumps is set via electro-optic modulation. We counted the repetition rates of two counter-propagating combs and performed an Allan deviation analysis. Up to 10~ms, the repetition rates are averaging down, meaning that coherent averaging can be performed up to this duration. On longer timescales, thermal drifts dominate, but we believe that the stability can be easily improved via a thermal stabilisation scheme based on the measurement of $\DFRep$.

% The stability of the pumps offset is inherent to the modulation frequency.

% =============================================
\begin{figure}[!h]
\centering
\includegraphics[width=.9\columnwidth]{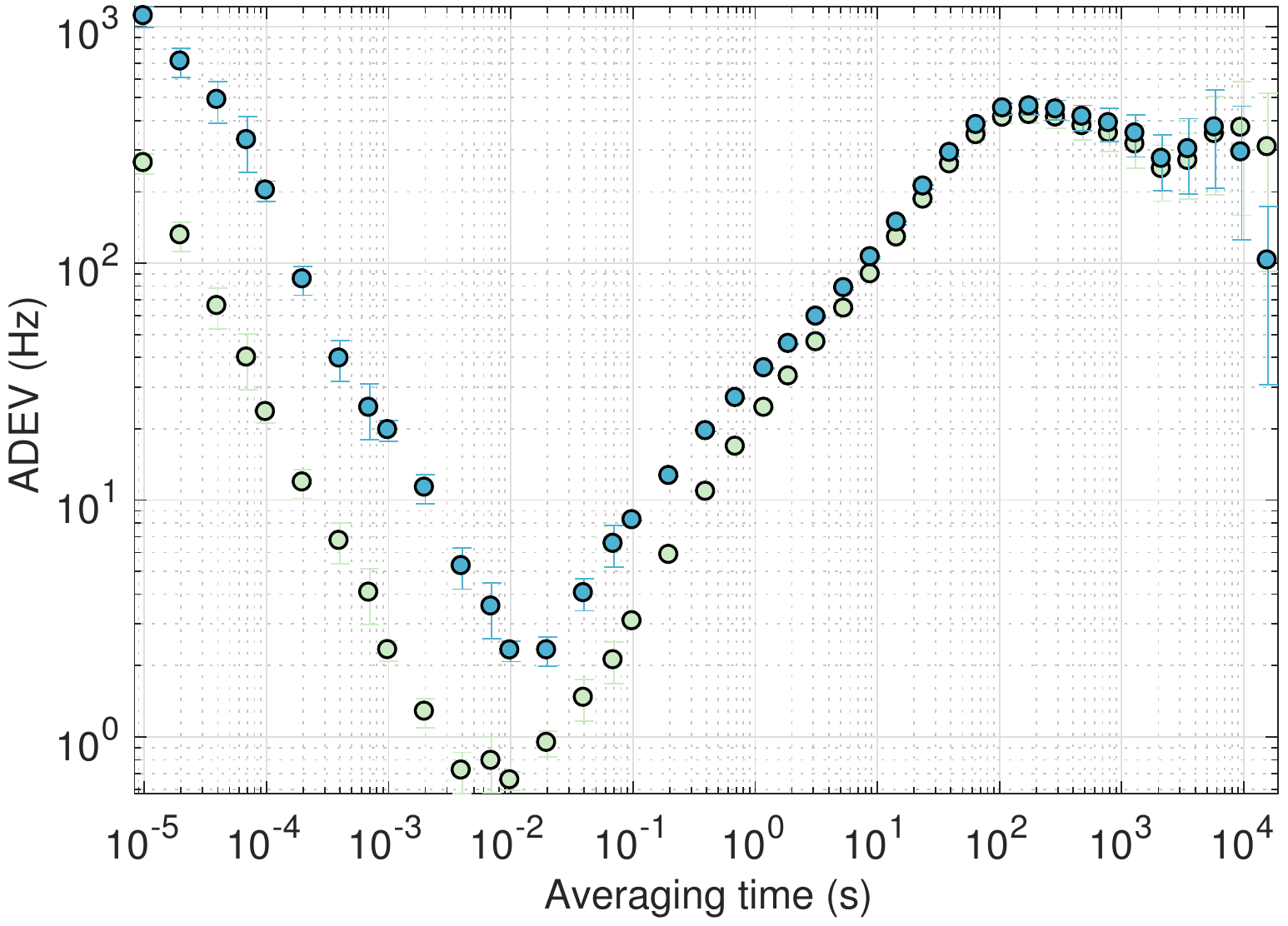}
\caption{
Overlapping Allan deviation of the repetition rates ($\sim 12$~GHz) of two counter-propagating combs. The marker colours corresponds to each repetition rate.
}\label{F7}
\end{figure}
% =============================================

% =============================================
\begin{figure*}[t]
\centering
\includegraphics[width=\textwidth]{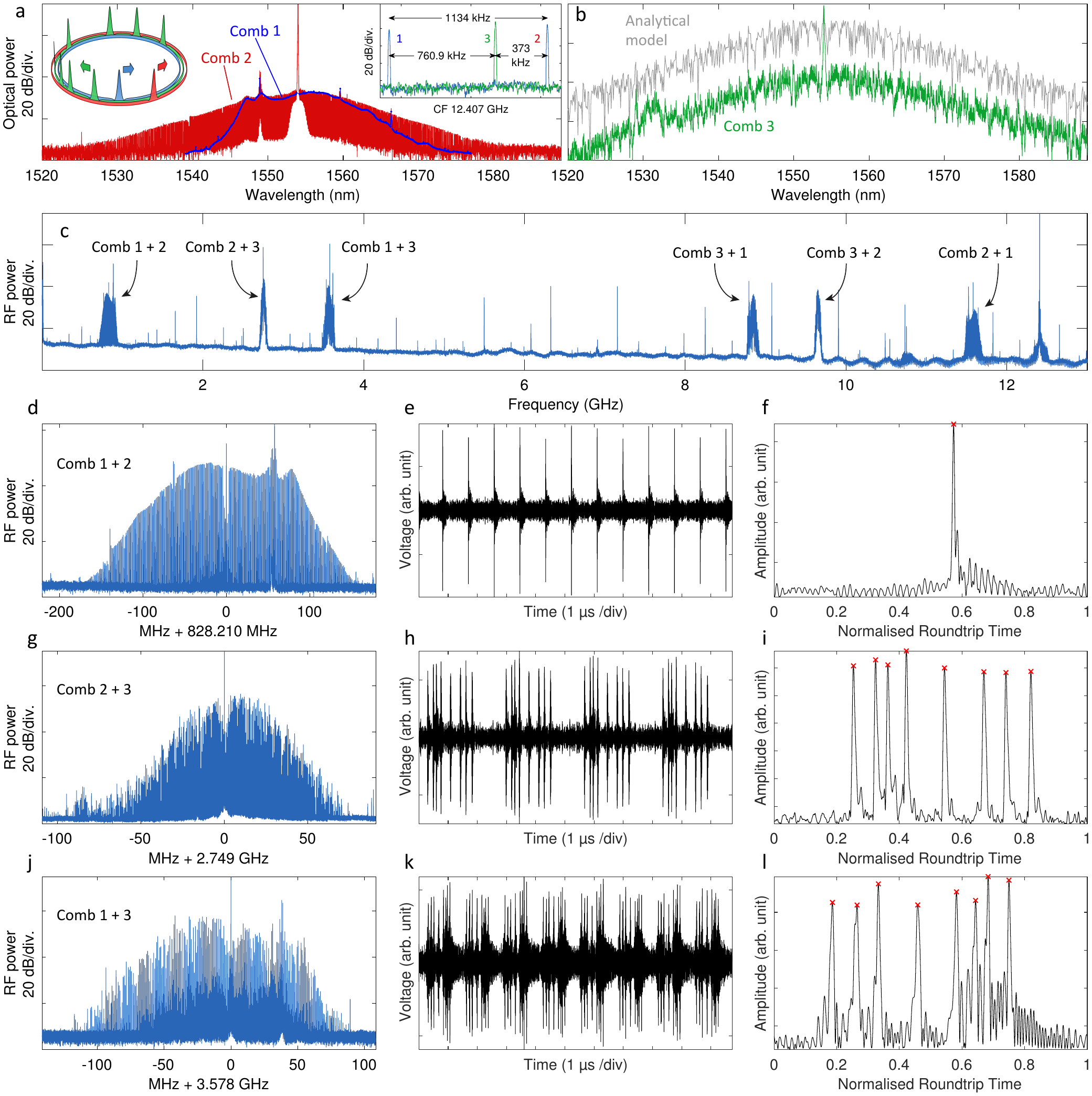}
\caption{
\textbf{Triple comb interferograms}
\textbf{(a)} Optical spectrum of the two CW co-propagating combs 1 and 2. The inset shows the three repetition rates of each comb. CF: centre frequency.
\textbf{(b)} Optical spectrum of the CCW comb 3 (green). The multi-soliton state spectrum is reconstructed after estimating the soliton number and position via dual comb imaging and using the analytical expression \eqref{eq:multisolSpectrum} (grey, the trace is shifted by + 20 dB for visualisation).
\textbf{(c)} Broadband spectrum of the triple comb interferogram. The spurious peaks in-between the RF combs arise due to the presence of modulation harmonics in the SSB modulated light signal that beat with the comb lines.
\textbf{(d,g,j)} Zoom in of each dual comb RF spectrum
\textbf{(e,h,k)} Time domain view of each interferogram (after selection with a bandpass filter).
\textbf{(f,i,l)} Envelope of the interferogram over one period, showing the convolution between the two selected pulse trains.
}\label{F11}
\end{figure*}
% =============================================

\smallskip\noindent\textbf{Triple comb interferogram} ---
We illustrate here another triple comb state and show time-domain-based measurement of a triple comb interferogram.
We employ the counter-propagating configuration, and set the modulation frequencies to $f_m = 2.75$~GHz and $f_m' = 3.58$~GHz on the modulator (\fref{F5}). In that case, the CW combs (1 and 2) correspond to single-soliton states, while the CCW comb 3 results from multiple solitons and features a complex modulation of the spectral envelope (\fref{F11}a,b). The soliton comb 1 is heavily impacted by a modal crossing on the short wavelength side, which decreases its bandwidth.
Heterodyning the combs creates a set of three RF combs centred at $f_m$, $f_m'$ and $|f_m'-f_m| = 828$~MHz and with a line spacing of 373~kHz, 761~kHz, and 1.13~MHz respectively (\fref{F11}d,g,j). The time domain interferogram was also acquired. The interferogram corresponding to heterodyning each pair of combs is retrieved after applying a bandpass filter to select the corresponding RF comb (\fref{F11}e,h,k). The envelope of each interferogram is also computed. One can note that the strong dispersive wave of comb 1 appears clearly in the modulated background in \fref{F11}f. Since two of the soliton states (1 and 2) contain a single soliton, they can be used to image the number of solitons and their relative position $\phi_i$ within a cavity roundtrip in the comb 3 (\fref{F11}f,i,j). In \fref{F11}i, it appears clearly that the comb 3 contains 8 solitons. To cross validate our position detection method, we compare the experimental optical spectrum of comb 3 with an analytical expression for $N = 8$ solitons with the relative positions $\phi_i/2\pi \in [0, 0.071, 0.110, 0.169, 0.291, 0.416, 0.487, 0.567]$, then the identical solitons circulating in the resonator produce a spectral interference on the single soliton spectrum~\cite{Brasch2015} following:
\begin{equation}
S^{(N)}(\mu) = S^{(1)}(\mu) \left( N + 2 \sum_{j\neq l} \cos \Big(\mu(\phi_j-\phi_l)\Big) \right). \label{eq:multisolSpectrum}
\end{equation}
Here $\phi_i\in[0,2\pi]$  is the position of the i-th pulse along the cavity roundtrip (assuming a roundtrip normalised to $2\pi$), $\mu$ is the comb mode index relative to the pump laser frequency and $S^{(1)}(\mu)$ is the spectral envelope of a single soliton following an approximate secant hyperbolic squared:
\def \bwModeN {\Delta \mu}
\begin{equation}
S^{(1)} \approx A \operatorname{sech}^2\left( \dfrac{\mu}{\bwModeN} \right),
\end{equation}
where $A$ is the power of the comb lines near the pump and $\bwModeN$ is the spectral width of the comb (in unit of comb lines). The expression \eqref{eq:multisolSpectrum} is computed with the retrieved soliton positions and the parameter $A$ and $\bwModeN$ are adjusted to fit the experimental comb amplitude and width. This analytical reconstruction is plotted on \fref{F11}b for comparison. The complex spectral interference pattern is faithfully reproduced, which validates the accuracy of our dual comb imaging technique. 

\putbib

\end{bibunit}

\end{document}